\documentclass[a4paper]{llncs}
\usepackage{latexsym} 
\usepackage{amssymb} 
\usepackage{amsmath} 
\usepackage{ntheorem}
\usepackage{mathtools}

\usepackage{amsmath}
\usepackage{soul} \usepackage{url}

\usepackage{lineno,xcolor}

\usepackage{tikz}
\usetikzlibrary{positioning}
\usetikzlibrary{arrows,snakes,backgrounds}

\def\mymedskip{\vskip\medskipamount}
\def\mymedbreak{\par \ifdim\lastskip<\medskipamount
  \removelastskip \penalty-100 \mymedskip \fi}
\def\myaftermedspace{\par \ifdim\lastskip<\medskipamount
  \removelastskip \penalty55\mymedskip\fi}
\newcommand{\eop}{{\unskip\nobreak\hfil\penalty50
          \hskip2em\hbox{}\nobreak\hfil$\Box$
          \parfillskip=0pt \finalhyphendemerits=0 \par}}
\newenvironment{proofn}[1]%
{\mymedbreak{\noindent\bf Proof #1.\enspace}}{\eop\myaftermedspace}
{\mymedbreak{\noindent\bf Proof of Theorem~\ref{#1}:\enspace}}
{\myaftermedspace}
\newtheorem{teor}{Theorem}
\newtheorem{defi}[teor]{Definition}
\newtheorem{fact}[teor]{Fact}
\newtheorem*{problemnn}{Problem}
\newtheorem{examp}[teor]{Example}
\newtheorem{lem}[teor]{Lemma}
\newtheorem{cor}[teor]{Corollary}
\newtheorem{con}[teor]{Conjecture}
\newtheorem{prop}[teor]{Proposition}
\newtheorem{rem}[teor]{Remark}
\newcommand{\beq}{\begin{equation}}
\newcommand{\eeq}{\end{equation}}
\newcommand{\beql}[1]{\begin{equation} \label{#1}}
\newcommand{\eeql}{\end{equation}}
\newcommand{\beqa}{\begin{eqnarray*}}
\newcommand{\eeqa}{\end{eqnarray*}}
\newcommand{\beqal}[1]{\begin{eqnarray} \label{#1}}
\newcommand{\eeqal}{\end{eqnarray}}
\newcommand{\beqan}{\begin{eqnarray}}
\newcommand{\eeqan}{\end{eqnarray}}
\newcommand{\bpf}{\begin{proof}}
\newcommand{\epf}{\end{proof}}
\newcommand{\bpfn}[1]{\begin{proofn}{#1}}
\newcommand{\epfn}{\end{proofn}}
\newcommand{\ben}{\begin{enumerate}}
\newcommand{\een}{\end{enumerate}}
\newcommand{\bit}{\begin{itemize}}
\newcommand{\eit}{\end{itemize}}

\newcommand{\bab}{\begin{abstract}}
\newcommand{\eab}{\end{abstract}}
\newcommand{\bke}{\begin{keywords}}
\newcommand{\eke}{\end{keywords}}

\newcommand{\btm}[1]{\begin{teor} \label{#1}}
\newcommand{\etm}{\end{teor}}
\newcommand{\btmn}[2]{\begin{teor}[#1] \label{#2}}
\newcommand{\etmn}{\end{teor}}
\newcommand{\ble}[1]{\begin{lem} \label{#1}}
\newcommand{\ele}{\end{lem}}
\newcommand{\bLe}[1]{\begin{Lemma} \label{#1}}
\newcommand{\eLe}{\end{Lemma}}
\newcommand{\blen}[2]{\begin{lem}[#1] \label{#2}}
\newcommand{\elen}{\end{lem}}
\newcommand{\bpn}[1]{\begin{prop} \label{#1}}
\newcommand{\epn}{\end{prop}}
\newcommand{\bex}[1]{\begin{examp} \label{#1}}
\newcommand{\eex}{\eop\end{examp}}
\newcommand{\bde}[1]{\begin{defi} \label{#1}}
\newcommand{\ede}{\end{defi}}
\newcommand{\bco}[1]{\begin{cor} \label{#1}}
\newcommand{\eco}{\end{cor}}
\newcommand{\bcorn}[2]{\begin{cor}[#1] \label{#1}}
\newcommand{\ecorn}{\end{cor}}
\newcommand{\bcon}[1]{\begin{con} \label{#1}}
\newcommand{\econ}{\end{con}}
\newcommand{\bfa}[1]{\begin{fact} \label{#1}}
\newcommand{\efa}{\end{fact}}
\newcommand{\bpr}[1]{\begin{problem} \label{#1}}
\newcommand{\epr}{\end{problem}}
\newcommand{\bprnn}[1]{\begin{problemnn} \label{#1}}
\newcommand{\eprnn}{\end{problemnn}}
\newcommand{\bprn}[2]{\begin{problem}[#1] \label{#2}}
\newcommand{\eprn}{\end{problem}}
\newcommand{\bexer}[1]{\begin{exercise} \label{#1}}
\newcommand{\eexer}{\end{exercise}}
\newcommand{\bre}[1]{\begin{rem} \label{#1}}
\newcommand{\ere}{\end{rem}}
\newcount\tpcnt
\newenvironment{tproblem}{%
  \global\advance\tpcnt1%
  \goodbreak\medskip\par\noindent\textbf{Problem~\the\tpcnt.}~}%
{%
  \goodbreak
}

\newenvironment{Solution}[1][]{%
  \goodbreak\smallskip\par\noindent\textbf{Solution{\if#1\empty\else~#1\fi}.}~}%
{%
  \goodbreak
}

\newcommand{\cB}{{\cal B}}

\newcommand{\cQ}{{\cal Q}}

\newcommand{\PG}{{\rm PG}}

\newcommand{\gd}{\delta}

\newcommand{\gre}{\epsilon}

\newcommand{\gl}{\lambda}

\newcommand{\Tm}[1]{Theorem~\protect\ref{#1}}

\newcommand{\Ta}[1]{Table~\protect\ref{#1}}

\newcommand{\Sec}[1]{Section~\protect\ref{#1}}

\newcommand{\bbF}{\mathbb{F}}

\newenvironment{hint}{\noindent {\bf Hint:} \enspace}{\eop\myaftermedspace}

\newenvironment{multisolution}[1]{\noindent {\bf Solution #1:} \enspace}{\eop\myaftermedspace}

\newcommand{\bqu}{\begin{question}}
\newcommand{\equ}{\end{question}}
\newcommand{\bs}{\begin{solution}}
\newcommand{\es}{\end{solution}}
\newcommand{\bh}{\begin{hint}}
\newcommand{\eh}{\end{hint}}
\newcommand{\bms}[1]{\begin{multisolution}{#1}}
\newcommand{\ems}{\end{multisolution}}

\newcommand{\btp}{\begin{tproblem}}
\newcommand{\etp}{\end{tproblem}}
\newcommand{\bts}{\begin{Solution}}
\newcommand{\ets}{\end{Solution}}

\newcounter{penumi}
\newenvironment{pit}{%
\begin{list}{(\roman{penumi})}{\usecounter{penumi}\setlength{\labelwidth}{1cm}\setlength{\itemindent}{0pt}\setlength{\topsep}{0pt}\setlength{\parsep}{0pt}\setlength{\partopsep}{0pt}\setlength{\itemsep}{0pt}}
} 
{\end{list}}
\newcommand{\bpit}{\begin{pit}}
\newcommand{\epit}{\end{pit}}

\newcommand{\choice}[5]{
\left\{ \begin{array}{ll} #1, & \mbox{#2};\\
                                   #3, & \mbox{#4}#5
\end{array}
\right. 
}

\begin{document} 

\title{
Optimal possibly nonlinear 3-PIR codes of small size} 

\author{Henk D.L.~Hollmann\inst{1} \and Urmas Luha\"a\"ar\inst{2}}
\institute{Institute of Computer Science, University of Tartu, Tartu 50409, Estonia\
\email{henk.d.l.hollmann@ut.ee}, ORCID https://orcid.org/0000-0003-4005-2369
\and
Institute of Mathematics and Statistics, University of Tartu, Tartu 50409, Estonia\
\email{urmas.luhaaar@ut.ee}}

\maketitle 

\begin{abstract}
First, we state a generalization of the minimum-distance bound for PIR codes. Then we
describe a construction for linear PIR codes using packing designs and use it to construct some new 5-PIR codes. Finally, we show that no encoder (linear or nonlinear) for the binary $r$-th order Hamming code produces a 3-PIR code except when $r=2$. We use these results to determine the smallest length of a binary (possibly nonlinear) 3-PIR code of combinatorial dimension up to~6. 
A binary 3-PIR code of length 11 and size $2^7$ is necessarily nonlinear, and we pose the existence of such a code as an open problem. 
\end{abstract} 


%
%


\section{\label{LSint}Introduction}

Private Information Retreval (PIR) scheme enables a user to extract a bit of information from a database, stored in encoded form on a multi-server distributed data storage system, without leaking information to the servers in which particular bit the user was interested in, see, e.g.,~\cite{chor95}.
A (binary) $t$-PIR code of length~$n$ and size $2^k$ is an encoder that encodes $k$ data bits one-to-one into $n$ encoded bits  in such a way that each data bit has $t$ mutually disjoint recovery sets. If the encoder employs only linear operations, then we speak of a  {\em linear\/} PIR code. Linear $t$-PIR codes can be used to implement a classical (linear) $t$-server PIR scheme \cite{chor95} with less storage overhead than the original scheme, by using the PIR code to emulate the $t$ servers \cite{fvy15}, \cite{fvy-arx15}; see also~\cite{var-you} for another explanation of how this magic is worked. 

A {\em batch code\/} is a special type of PIR code where for {\em any\/} batch of $t$ data symbols, there exist $t$ mutually disjoint recovery sets. 
Batch codes were initially introduced in \cite{Ishai04} as a method to improve load-balancing in distributed data storage systems. Later, so-called {\em switch codes\/} (a special case of batch codes) were proposed in~\cite{Wang2013} as a method to increase the throughput rate in network switches. In such applications, there is no need for the batch code to be linear.
We remark that a PIR or batch code can be nonlinear because the associated code is nonlinear, or because it consists of a nonlinear encoder onto a linear code.

For an overview of PIR- and batch-type codes and other similar codes, we refer to~\cite{ska2018}. 
In this paper, all PIR codes are multiset primitive \cite{ska2018}, and we will mostly consider only binary codes. Precise definitions will be given in the next section.

For linear PIR-codes, much work has been done to find bounds on the smallest~$n$ for which a linear $t$-PIR code of dimension~$k$ and length~$n$ exists, see for example \cite{ky21} for a recent overview. For linear batch codes, the situation is similar.
Nonlinear PIR-codes are interesting combinatorial objects in their own right, but in contrast, virtually nothing is known about their possible parameters. In fact, we do not know a single example of an ``interesting'' nonlinear PIR code, that is, with parameters for which no linear PIR-code exists. One of our aims in this paper is to at least identify some parameters for which such an interesting nonlinear code could exist, were we concentrate on 3-PIR codes since there are linear optimal $t$-PIR codes for $t=1,2$ (see, e.g., \cite[page 560]{ky21}).

The contents of this paper are as follows. In \Sec{LSpir}, we define the notion of a $t$-PIR code and various other notions that we will need. Our results strongly depend on a simple bound on the minimum distance of a (linear or nonlinear) $t$-PIR code. In \Sec{LSmdb}, we derive a generalization of this lower bound for a broad class of (not necessarily linear) PIR-like codes. For linear 3-PIR and 3-batch codes, the optimal codes are known.  Bounds and constructions for linear 3-PIR codes and some generalizations of these constructions are discussed in~\Sec{LSoldbounds}. In \Sec{LSham} we prove one of our main results, stating that no encoder for a binary length $2^r-1$ Hamming code with $r\geq 3$ is a 3-PIR code.
We use this result to determine the optimal length of 3-PIR codes of size~$2^k$ for $1\leq k\leq 6$ in \Sec{LS3pirex}, and we pose the question of the existence of a (necessarily nonlinear) 3-PIR code of length~$11$ and size~$2^7$ as an open problem. We end with some conclusions and further questions in \Sec{LScon}.
\section{\label{LSpir}Preliminaries}
%
%
Let~$q$ be a positive integer. We use $\Sigma$ to denote an alphabet with~$q$ symbols; if $q$ is a prime-power, we identify these symbols with the $q$ elements of the finite field~$\bbF_q$ of size~$q$.  For a positive integer $n$, we let $[n]$ denote the set $\{1, \ldots, n\}$, and we use this set to index the positions in code words of length~$n$. 

Informally, PIR- and batch-type codes are characterized by the property that given the encoded data, certain {\em simultaneous\/} requests for specific data symbols can each be handled by reading and decoding data from a set of positions called a {\em recovery set\/}, where these sets are supposed to be of bounded size, with limited overlap between the sets. 
We now introduce some useful terminology to make this precise.

\bde{LDenc}\rm
A 
{\em $k$-to-$n$ encoder\/} over an alphabet~$\Sigma$ is a one-to-one map $\gre: \Sigma^k\rightarrow \Sigma^n$; the image~$C=C_\gre$ of~$\gre$ is referred to as the {\em associated code\/} of~$\gre$. 
By definition, such an encoder $\gre$ has a {\em decoder\/}~$\gd: C\rightarrow \Sigma^k$ with the property that if $c=\gre(a)$, then $\gd(c)=a$. We refer to $\gre$ as a $q$-ary encoder if $|\Sigma|=q$. 
\ede
Let  $I=\{i_1, \ldots, i_s\}\subseteq [n]$ with $i_1<\ldots  <i_s$. Given a code word $c\in\Sigma^n$, the {\em restriction\/} $c_I$ of $c$ to $I$ is the word $c_I=(c_{i_1}, \ldots, c_{i_s})$. 
\bde{LDrec}\rm We say that $I$ is a {\em recovery set\/} of the $j$-th data symbol for a $k$-to-$n$ encoder~$\gre$ over~$\Sigma$ if for every $a\in \Sigma^k$, when $c=\gre(a)$, the restriction $c_I$ of $c$ to~$I$ uniquely determines~$a_j$; it is called {\em minimal\/} 
if no proper subset of~$I$ 
has this property.

%
A {\em query\/} of~$\gre$ is a sequence $i_1, \ldots, i_t$ of (not necessarily distinct) elements of~$[n]$. Given a code word  $c=\gre(a)$,  the query $i_1, \ldots, i_t$ should be considered as a request to obtain the data symbols $a_{i_1}, \ldots, a_{i_t}$. We will say that the sets $I_1, \ldots, I_t\subseteq [n]$ {\em serve\/} the query of~$\gre$ if for every $j\in [t]$, the set $I_j$ is a recovery set of $\gre$ for the $i_j$-th data symbol. We say that $I_1, \ldots, I_t$ serve the query with {\em width\/}~$w$ and {\em multiplicity\/}~$\mu$ if $|I_j|\leq w$ ($j=1, \ldots, t$) and if every position~$i\in [n]$ occurs in at most $\mu$ of the sets $I_1, \ldots, I_t$.
\ede
%
%
Now we are ready for a definition of batch-type codes. 
\bde{LDRbatch}\rm Let $\gre$
be a 
$k$-to-$n$ encoder over~$\Sigma$, let $w,\mu$ be positive integers, and let $\cQ$ be a collection of queries of~$\gre$. We say that $\gre$ 
is a {\em $(\cQ,w,\mu)$-batch code\/} if $\gre$ can serve every query in~$\cQ$ with width at most~$w$ and multiplicity at most~$\mu$. The encoder $\gre$ is a $(t,w,\mu)$-PIR code if $\gre$ is a $(\cQ, w,\mu)$-batch code with $\cQ$ consisting of all queries of the form $i, i, \ldots, i$ ($t$ times) with $i\in [k]$; a $(t,\infty,1)$-PIR code is called a {\em  $t$-PIR code\/}. The encoder $\gre$ is  a {\em $t$-batch code\/}  if $\gre$ is a $(\cQ, \infty, 1)$-batch code with $\cQ$ consisting of all queries of the form $i_1, \ldots, i_t$ with $i_1, \ldots, i_t\in [k]$.
\ede
%

More informally, a recovery set for a data symbol allows the recovery of a certain data symbol by inspecting only the code word symbols in the positions of the recovery set. 
Then a $t$-PIR code has the property that every encoded data symbol has $t$ mutually disjoint recovery sets, while for a $t$-batch code we can find $t$ mutually disjoint recovery sets for every batch of $t$ data symbols.

We remark that what we call here a batch code is referred to by some authors as a primitive (multiset) batch code, see, e.g., \cite{ska2018}.

A {\em linear\/} $k$-to-$n$ encoder over a $q$-ary alphabet is an $\bbF_q$-linear map $\gre: \bbF_q^k\rightarrow \bbF_q^n$, which can thus be represented by a $k\times n$ matrix~$G$ over~$\bbF_q$; here $G$ is the generator matrix of the associated linear code~$C=\gre(\bbF_q^k)$. In this case, a set $I\subseteq [n]$ is a recovery set for the $j$-th data symbol if and only if some $\bbF_q$- linear combination of the columns of~$G$ indexed by $I$ sum up to $e_j$, the $j$-th unit vector in~$\bbF_q^k$, for a proof see  \cite[Theorem 1]{ls14}.  

In this paper, we are mainly interested in ``optimal'' binary $t$-PIR and $t$-batch codes with $1\leq t\leq 4$. 
\bde{LDopt}\rm Let $k$ and $t$ be positive integers. 
We let $P(k,t)$, $PL(k,t)$,
$B(k,t)$, and $BL(k,t)$
denote the smallest length~$n$ of a binary possibly nonlinear $t$-PIR code, a binary linear $t$-PIR code, 
a binary possibly nonlinear $t$-batch code, or a binary linear $t$-batch code, of size~$2^k$, respectively.
\ede  
We will refer to a code of the above types with an optimal, minimal length as an {\em optimal\/} code for that type.
\section{\label{LSmdb}The minimum-distance bound for batch-type codes}
Let $\Sigma$ denote an alphabet of size~$q$. An $(n,M,d)_q$-code $C$ is a subset of~$\Sigma^n$, of size~$M$, where any two distinct code words in~$C$ have (Hamming) distance at least~$d$. Here, the (Hamming) distance between two words $v,w\in \Sigma^n$ is the number of positions in which $v$ and $w$ differ. An $[n,k,d]_q$ code is a {\em linear\/} code of length $n$ and dimension~$k$ over $\bbF_q$, with minimum distance~$d$.
One of the very few known lower bounds for the length of a $t$-PIR code of a given size results from the observation that a $t$-PIR code must have minimum distance at least~$t$. This was first stated for binary linear batch codes in~\cite{ls14} and for non-linear batch codes over general alphabets in~\cite{zs15}. See also \cite{ska2018}, \cite{zs16}, and~\cite{lr18,lr18-arx} where the result was stated for PIR codes.
Here we present a slight generalization of these results.
\btm{LTmindistbound}\rm 
Let $C$ be an $(n,q^k, d)_q$-code over an alphabet~$\Sigma$, and suppose
that~$C$ has an encoder $\gre: \Sigma^k\rightarrow C$ 
that is  a $(t,\infty,\mu)$-PIR code. Then $\lceil t/\mu\rceil\leq d$. 
\etm
\bpf
Let $\gd: C\rightarrow \Sigma^k$ be the corresponding decoder.
Let $c^{(1)}, c^{(2)}$ be distinct code words from~$C$. Then there is an $s$ such that $\gd(c^{(1)})_s\neq \gd(c^{(2)})_s$. By our assumption on~$C$, there are sets $I_1, \ldots, I_t$ that serve the query 
$s, s, \ldots, s$ ($t$ times) with multiplicity at most~$\mu$. So for every position set $I_j$, the restrictions $c^{(1)}_{I_j}$ and $c^{(2)}_{I_j}$ determine distinct data symbols,  hence $I_j$ must contain a position $i_j$ for which $c^{(1)}_{i_j}\neq c^{(1)}_{i_j}$.
By the multiplicity condition there must be at least $\lceil t/\mu\rceil$ distinct positions among $i_1, \ldots, i_t$, so as a consequence, $c^{(1)}$ and $c^{(2)}$ differ in at least $\lceil t/\mu\rceil$ positions. Since the code words were arbitrary, we conclude that $d\geq \lceil t/\mu\rceil$.
\epf
We will refer to a code that attains the bound in~\Tm{LTmindistbound} as {\em distance-optimal\/}.
\section{\label{LSoldbounds}Some bounds and constructions}
%
For later use, we first state the following simple result.
\btm{LTsimple}\rm If $P(k,2t-1)=PL(k,2t-1)$, then $P(k,2t)=PL(k,2t)=P(k,2t-1)+1$.
\etm
\bpf
Suppose that the condition in the theorem holds, and let $C$ be a linear $(2t-1)$-PIR code of dimension~$k$ and length $n=P(k,2t-1)$. Then by a well-known argument (see \cite{fvy-arx15}), the extended code $\overline{C}$ (adding an overall parity-check bit)  is a $(2t)$-PIR code, hence $P(k,2t)\leq P(k,2t-1)+1$. On the other hand, if $C'$ is any $s$-PIR code of size~$2^k$ and length~$n$, then the code obtained from~$C'$ by deleting a position is obviously an $(s-1)$-PIR code. By taking $s=2t$, we conclude that $P(k,2t-1)\leq P(k,2t)-1$. Combining these inequalities shows that $P(k,2t)=P(k,2t-1)+1$, and since $\overline{C}$ has length~$P(k,2t-1)+1=P(k,2t)$, we also have that $P(k,2t)=PL(k,2t)$.
\epf
As a consequence of \Tm{LTsimple}, we can restrict our search for binary nonlinear $t$-PIR codes to the cases where $t$ is odd. We obviously have $P(k,1)=PL(k,1)=k$ and $P(k,2)=PL(k,2)=k+1$, where the optimal codes are the entire $k$-dimensional space and the even-weight vectors in a $(k+1)$-dimensional space, respectively (see, e.g., \cite[page 560]{ky21}). This leads us to consider the case where $t=3$.

In \cite{rv17}, it was shown that a linear 3-PIR code with dimension~$k$ and length~$n$, so with redundancy $r=n-k$, satisfies the bound $r(r-1)/2\geq k$. Moreover, this bound is attained by the codes with generator matrix of the form $(I_k P)$, where $P$ is the $k\times r$ matrix that has rows consisting of distinct binary vectors of weight 2 (note that such a matrix exists by the condition on $k$ and~$r$). We even have the following.
\btm{LT3pir}\rm Let $k\geq1$ be integer. The code $C$ with generator matrix $(I_k P)$ as defined above is 3-batch, and the extended code is 4-batch. Hence both are optimal linear codes, $BL(k,3)=PL(k,3)$, and $BL(k,4)=PL(k,4)$. Both the code $C$ and its extension are also distance-optimal.
\etm
\bpf
The batch properties of the two codes can easily be proved directly, but also follow from~\cite[Lemma 3, 4, 5]{vy16} since the matrices of the form $(I_k P)$ as defined above are systematic. Since $PL(k,4)=PL(k,3)+1$ (see \cite{fvy-arx15}), both codes must be optimal both as PIR and as batch codes. Since the code~$C$ has code words of weight 3 in its generator, by the minimum distance bound \Tm{LTmindistbound}, it has distance 3, and the extension has minimum distance 4.
\epf
In fact, the above code construction can be generalized. To this end, we need a special type of combinatorial structure. Let $v\geq k\geq t$. A {\em $t-(v,k,\gl)$ packing design\/} or, more briefly, a {\em packing\/}, consists of a collection~$\cB$ of subsets of $[v]$, each of size~$k$, with the property that any subset of~$[v]$ of size~$t$ occurs in at most~$\gl$ sets in~$\cB$. We will refer to the elements of~$[v]$ as {\em points\/} and to the elements of~$\cB$ as {\em blocks\/}. We write $D_\gl(v,k,t)$ to denote the {\em packing number\/}, the largest possible number of blocks in a $t-(v,k,\gl)$ packing; in the case where $\gl=1$, we denote the packing number by~$D(v,k,t)$. For a general overview of packing designs, we refer to 
\cite[Part IV, Section 40]{cd07}.

Here, we will be interested in the case $t=2$ and $\gl=1$. Note that in this case, any two blocks of the design intersect in at most one point (indeed, otherwise a pair of points from the intersection would be contained in at least two blocks).  
We now have the following generalization of~\Tm{LT3pir}.
\btm{LTtpir}\rm 
Let $r,t$ be positive integers with $r\geq t-1$, and let $k$ be a positive integer such that $k\leq D(r,t-1,2)$. Let $P$ be a $k\times r$ matrix whose rows are the incidence vectors of~$k$ pairwise distinct blocks from a $2-(r,t-1,1)$ packing design with at least $k$ blocks (note that this is possible by the condition on~$k$). Then the matrix $(I_k P)$ is the generator matrix of a $t$-PIR code. As a consequence, we have that $PL(k,t)\leq k+r$, where $r$ is the smallest integer for which $k\leq D(r,t-1,2)$.
\etm
\bpf
By the properties of a packing design, this follows immediately from~\cite[Lemma 7]{fvy-arx15} or~\cite[Lemma 7]{fvy15}.
\epf
Strictly speaking, the above result is not new. But the authors of~\cite{fvy15} did not explicitly make the connection with packing designs, so they did not quantify their result except for the case of Steiner systems. 

Note that this theorem indeed generalizes \Tm{LT3pir} since in the case where $t=3$, a $2-(r,2,1)$ packing design is simply a collection of pairs from $[r]$, so that $D(r,2,2)=r(r-1)/2$. Since, as remarked before, $PL(k,4)=PL(k,3)+1$, the next interesting case of the above theorem is when $t=5$. Interestingly, the packing numbers $D(r,4,2)$ are completely known. 
\btm{LTpack4}\rm {\bf (See \cite{brou79})}
Let
\[U(r,4,2)=\left\lfloor \frac{r}{4}\left\lfloor\frac{r-1}{3}\right\rfloor\right\rfloor,\]
and write
\beql{LEpack} J(r,4,2) =\choice{
U(r,4,2)-1
}{for $r\equiv 7$ or $10 \pmod {12}$}{
U(r,4,2)
}{otherwise}{.}
\eeql
Then $D(r,4,2)=J(2,4,r)$ if $r\notin\{8,9,10,11,17,19\}$ and $D(r,4,2)=J(r,4,2)-\gre$ with $\gre=1$ for $r\in \{9,10,17\}$ and $\gre=2$ for $r\in \{8,11,19\}$.
\etm
%
%
In the next example,we discuss some applications of~\Tm{LTtpir} and~\Tm{LTpack4}.
\bex{LEpirpack4}\rm 
We mention some improvements of \cite[Table III]{fvy-arx15}. \\
(i) First, $D(12,4,2)=9$, so $P(9,5)\leq9+12=21$ and $P(9,6)\leq22$, which improves the known value by 1, but loses against the more recent \cite[Table 1]{ky21}. \\
(ii) We have $D(15,4,2)=15$ and $D(16,4,2)=20$. So $P(15,5)\leq 30$, hence $P(15,6)\leq 31$, improving the value in \cite[Table III]{fvy-arx15} by 3, and $P(16+i,5)\leq32+i$, hence  $P(16+i,6)\leq 33+i$, for $i=0, \ldots,4$, improving the values in~\cite[Table III]{fvy-arx15} by 4. These results are similar to those in \cite{GST21} (unpublished).
\eex
%
%

%
\section{\label{LSham}The Hamming codes as PIR-codes}
For an integer $r\geq2$, the binary $r$-th order Hamming code is a linear code of length~$n=2^r-1$ and dimension~$k=2^r-1-r$, with the $k\times n$  parity-check matrix~$H_k$ whose columns are the nonzero binary vectors of length~$r$.  Obviously, these codes have minimum Hamming distance 3. We will now prove the following.
\btm{LTham}\rm
For $r\geq2$, the all-one word 1 is in the $r$-th order Hamming code. Moreover, let $r\geq3$ and
suppose that for some encoder for the $r$-th order  Hamming code, the position subsets $I_1, I_2, I_3$ are three mutually disjoint, minimal recovery sets for a particular data bit. Then for every code word~$c$, both $c$ and its complement $1+c$ decode to the same value of that data bit. 
\etm
\bpf
It is natural to label the positions with the nonzero binary vectors of length~$r$.
In what follows, we will not distinguish between a set $S\subseteq \bbF_2^r\setminus \{0\}$ and its characteristic vector $\chi_S$ of length~$2^r-1$ that has a 1 in the positions of~$S$ and a 0 in the other positions. Note that with this convention, a set $S=\{v,w,v+w\}$ corresponds to a word of (minimal) weight 3 in the Hamming code, so the minimum weight vectors in the Hamming code correspond to the lines in the projective geometry $\PG(r-1,2)$.  Note also that every point in~$\PG(r-1,2)$ is on $(2^r-2)/2=2^{r-1}-1$ lines, so for $r\geq 2$ the all-one vector 1 is contained in the code. In what follows, we associate the points of~$\PG(r-1,2)$ with the nonzero vectors in~$\bbF_2^r$.

First, we claim that a line $L$ intersecting two of the sets $I_1, I_2, I_3$ also intersects the third one. Indeed, if not, we may assume without loss of generality that $L$ intersects $I_1$ only in $P$ and does not intersect $I_3$. Let $\ell$ be the code word corresponding to the line~$L$. Then for every code word $c$, the code words $c$ and $c+\ell$ have the same restriction to $I_3$, so decode to the same value for the data bit, while their restrictions to $I_1$ differ exactly in position~$P$. As a consequence, the restriction of $c$ to $I_1\setminus \{P\}$ already contains sufficient information to decode, contradicting the minimality of~$I_1$. 

Next, we claim that none of $I_1, I_2, I_3$ contains a line. Indeed, suppose that $I_1$ contains the line $L=\{P_1, P_2, P_3\}$. Let $R$ be a point in~$I_3$. Then $L$ and $R$ together span a $\PG(2,2)$. Now consider the lines $L_i$ though $P_i$ and $R$ ($i=1,2,3$). By the first claim, the third point~$Q_i$ on the line containing~$R$ and~$P_i$ is in~$I_2$.  Then the third line through $P_1$ in this $\PG(2,2)$ is $\{P_1, Q_1, Q_2\}$, intersecting $I_1$ in one point and $I_2$ in two points, contradicting the first claim.

Finally, as a consequence of the above two claims, if $P,Q$ are two points in some $I_i$, then the third point $R$ on the line $L$ through $P$ and $Q$ is outside $I_i$ and by the first claim $R$ is outside $I_1\cup I_2\cup I_3$. Consider any code word $c$. If $\ell$ is the code word corresponding to the line $L$, then since $c$ and $c+\ell$ have the same restriction to the sets $I_j$ with $j\neq i$, they decode to the same value of the data bit. Since the two points and the set $I_i$ are arbitrary, it follows that on each of $I_1, I_2, I_3$, the restrictions that have even weight all decode to the same value of the data bit, and the restrictions that have odd weight all decode to the complement of that value.  

Since the all-one word is contained in the code, it follows from the above that to prove the theorem, we are done if we can show that each of the sets $I_1, I_2, I_3$ has even size. To this end, let~$H$ consist of the all-zero vector 0 together with all the nonzero vectors associated with the points outside $I_1\cup I_2\cup I_3$. By the minimality of the $I_i$'s, no line containing two points from $H\setminus \{0\}$ can have its third point outside $H$, hence $H$ is a subspace of~$\bbF_2^r$. Moreover, for every $i$, the line through two points on~$I_i$ has its third point on~$H$, hence $I_i$ is contained in a coset of~$H$. Moreover, by our first claim, each of these cosets are distinct, and since $H$ and the $I_i$ together partition~$\bbF_2^r$, we conclude that $|H|=|I_1|=|I_2|=|I_3|=2^{r-2}$. As a consequence, for every $i$, the set $I_i$ indeed has even size provided that $r\geq3$. 
\epf
Obviously, since the all-one vector is a code word, \Tm{LTham} implies that no encoder for the the $r$-th order Hamming code with $r\geq3$ can be a 3-PIR code. Since the second order Hamming code is just the repetition code of length 3, which is easily seen to be a linear 3-PIR code, we have proved the following. 
\bco{LCham}\rm
The $r$-th order Hamming code ($r\geq 2$) has a (linear or nonlinear) 3-PIR encoder if and only if $r=2$. 
\eco
\section{\label{LS3pirex}Optimal (not necessarily linear) 3-PIR codes}
Earlier, we have already remarked that the best $q$-ary (not necessarily linear) 1-PIR code of size $q^k$ has length $n=k$ and consists of all words of length~$k$, and the best 2-PIR code of size $q^k$ has length $n=k+1$ and consists of all words $c=(c_0, \ldots,c_{n-1})$ for which $\sum c_i =0$ (in the binary case, this is the even-weight code).

In~\Tm{LT3pir} we have seen that a binary linear 3-PIR code of length~$n$ and dimension~$k$, so with a linear encoder and completely described by a $k\times n$ generator matrix, has a redundancy $r=n-k$ satisfying $r(r-1)/2\geq k$. We also saw that codes satisfying this bound exist: they have a generator matrix of the form $G=(I_k P)$ where $P$ is a $k\times r$ matrix that has distinct weight-two vectors as its rows. In \Ta{LTlin3pir} below, we list the optimal length of a binary linear $k$-dimensional 3-PIR code of this form, for various values of~$k$.
\begin{table}[htb]
\begin{center}
\begin{tabular}{|c|ccc ccc cc|}
\hline
$k$& 1&2&3&4&5&6&7&8\\ \hline
$n$& 3&5&6&8&9&10&12&13\\
\hline
\end{tabular}
\end{center}
\caption{\label{LTlin3pir}Optimal (smallest) length of binary linear $k$-dimensional 3-PIR codes}
\end{table}

\noindent
{\em A priory\/}, it is possible that there exist shorter non-linear codes. By the minimum-distance bound in~\Tm{LTmindistbound}, any 3-PIR code has minimum distance $d\geq3$. In \Ta{LTAnd} we list the values of $A_2(n,3)$, the maximum number $M$ of code words in a binary code of length~$n$ and distance~3, see~\cite{br-tab}. 
%
\begin{table}[htb]
\begin{center}
\begin{tabular}{|c|c ccc cc cccc|}
\hline
$n$& 3&4&5&6&7&8&9&10&11&12\\ \hline
$M$& 2&2&4&8&16&20&40&72&144&256\\
\hline
\end{tabular}
\end{center}
\caption{\label{LTAnd}Maximum size $A_2(n,3)$ of a binary code of length $n$ and minimum distance~3}
\end{table}
Inspection of \Ta{LTAnd} shows that there are no shorter binary codes of length~$n$ and minimum distance~3 than those in \Ta{LTlin3pir} for $k=1,2,3,5,6$. For $k=4$, there is a unique code of length 7, size 16, and minimum distance 3 (see \cite{za52}), which is the Hamming code of that length. We have shown that there is no encoder (linear or nonlinear) that turns that code into a 3-PIR code. For $k=7$, there are 7398 inequivalent binary codes of length 11, size 144, and minimum distance 3 (see \cite{os99}). As a consequence, there are many nonlinear binary codes of length 11, size $2^7$ and minimum distance 3. We do not know if there exist a (nonlinear) 3-PIR code with these parameters.
\bprnn{LP11}\rm
Does there exist a (nonlinear) binary 3-PIR code of length 11 and size $2^7$?
\eprnn
In fact, we believe that the answer is no. 
%
Indeed, we suspect that $P(k,3)=PL(k,3)$, that is, for every $k\geq1$, there are no nonlinear codes of size~$2^k$ with a shorter length than the linear 3-PIR codes of size~$2^k$ in~\Tm{LT3pir}, but presently we have neither a proof nor a counterexample.
\section{\label{LScon}Conclusions}
First, we have shown how packing designs can be used to construct new PIR codes. 
Then, we have shown that for $r\geq2$, the $r$-th order Hamming code has a (linear or nonlinear) 3-PIR encoder if and only if $r=2$. Using the fact that a (linear or nonlinear) $t$-PIR code has minimum Hamming distance at least~$t$,  this result has allowed us to determine $P(k,3)$,  the shortest length of a (not necessarily linear) 3-PIR code of size $2^k$, for $k\leq 6$. We posed the existence of a (necessarily nonlinear) 3-PIR code of length 11 and size $2^7$ as an open problem.

%
\section*{Acknowledgments}
The research of the first author was supported by the Estonian Research Council grant PRG49. It is a great pleasure to thank our colleagues Vitaly Skachek, Karan Khathuria, and Ago-Erik Riet for their help in preparing this paper.
%
\bibliographystyle{abbrv} 
%

%

\end{document}